\documentclass{ws-procs9x6}
\usepackage{amssymb}

\usepackage{amsmath}

\input{tcilatex}

\begin{document}

\title{GLUEBALL CORRELATORS AS HOLOGRAMS}
\author{HILMAR FORKEL}

\address{Institut f{\"u}r Physik, Humboldt-Universit{\"a}t zu Berlin,\\
D-12489 Berlin, Germany \\
and \\
Departamento de F{\'i}sica, ITA-CTA, \\
12.228-900 S{\~a}o Jos{\'e} dos Campos, S{\~a}o Paulo, Brazil\\
E-mail: forkel@ift.unesp.br}

\begin{abstract}
We investigate the dynamical content of both hard- and soft-wall
approximations to holographic QCD by deriving the corresponding glueball
correlation functions and by confronting them with a variety of QCD results.
We further calculate the glueball decay constants in both holographic duals,
discuss emerging limitations and improvement strategies, and comment on a
recent attempt to generalize the glueball correlator in the soft-wall
background.
\end{abstract}

\bodymatter

\section{Introduction}

The discovery and ongoing development of gauge/string dualities \cite{aha00}
has opened up a new and exciting frontier for nonperturbative QCD. Already
the current, first generation of ``holographic QCD'' or ``AdS/QCD'' duals
with their bold approximations is beginning to provide new and often
surprising analytical insights into the elusive infrared (IR) sector of the
strong interactions.

By now a rather large set of static hadron properties has been calculated in
the AdS/QCD framework (for recent reviews see e.g. Ref. [2]). The majority
of this work was based on the two currently most popular dual candidates,
i.e. the hard- \cite{pol02} and soft-wall \cite{kar06} backgrounds. Since
even these rather minimal gravity backgrounds turn out to describe most
calculated static hadron properties at an astonishing 10-30\% accuracy
level, it becomes increasingly important to explore their capacity and
limitations in describing more detailed and sensitive QCD amplitudes. One
such class of amplitudes comprises the hadron form-factors, and several of
them have already been estimated holographically \cite{ff}. Another
important set of hadron amplitudes are the $n$-point functions of hadronic
interpolating fields, and among these the two-point correlators play a
special role, not least because detailed QCD results are available for most
of them.

We have therefore recently advocated \cite{for08} to put AdS/QCD dual
candidates to more stringent tests by evaluating their predictions for
hadron correlators and by confronting those with QCD information from the
lattice \cite{che06}, the operator product expansion (OPE) including hard
instanton contributions to the Wilson coefficients \cite{for01}, a
hypothetical UV gluon mass suggested to encode the short-distance behavior
of the static quark-antiquark potential \cite{che99}, and a scaling
low-energy theorem \cite{let} based on the trace anomaly. (AdS/QCD
correlators were recently also studied in Refs. [11--13].) In the following
we will outline the main steps of implementing this program in the scalar
glueball channel \cite{for08}. To this end, we review the calculation of the
scalar glueball spectra, decay constants and correlators in both hard- and
soft-wall backgrounds, and we comment on a recent attempt \cite{col07} to
generalize the soft-wall correlator.

\section{Glueball spectra and decay constants}

The holographic hard- and dilaton soft-wall duals are both based on
five-dimensional bulk geometries of ``Poincar\'{e} domain wall'' type%
\begin{equation}
ds^{2}=g_{MN}\left( x\right) dx^{M}dx^{N}=e^{2A\left( z\right) }\frac{R^{2}}{%
z^{2}}\left( \eta _{\mu \nu }dx^{\mu }dx^{\nu }-dz^{2}\right)  \label{metric}
\end{equation}%
where $\eta _{\mu \nu }$ is the four-dimensional Minkowski metric and
conformal invariance of the dual gauge theory in the UV requires $A(z)%
\overset{z\rightarrow 0}{\longrightarrow }0$. The soft wall additionally
contains a bulk dilaton field $\Phi \left( z\right) $.

Scalar QCD glueballs are interpolated by the lowest-dimensional gluonic
operator carrying vacuum quantum numbers, 
\begin{equation}
\mathcal{O}_{S}\left( x\right) =G_{\mu \nu }^{a}\left( x\right) G^{a,\mu \nu
}\left( x\right)
\end{equation}%
(where $G_{\mu \nu }^{a}$ is the gluon field strength). Since $\mathcal{O}%
_{S}$ has conformal dimension $\Delta =4$ (at the classical level), the
AdS/CFT dictionary \cite{aha00} prescribes its dual string modes $\varphi
\left( x,z\right) $ to be the normalizable solutions of the scalar wave
equation in the bulk geometry (\ref{metric}) (and possibly other background
fields) with the UV behavior $\varphi \left( x,z\right) \overset{%
z\rightarrow 0}{\longrightarrow }z^{\Delta }\phi \left( x\right) $. The
latter implies that the square mass $m_{5}^{2}R^{2}=\Delta \left( \Delta
-d\right) $ of the bulk field $\varphi $ vanishes for $d=4$, so that its
minimal action takes the form%
\begin{equation}
S\left[ \varphi ;g,\Phi \right] =\frac{1}{2\kappa ^{2}}\int d^{5}x\sqrt{%
\left| g\right| }e^{-\Phi }g^{MN}\partial _{M}\varphi \partial _{N}\varphi .
\label{sd}
\end{equation}

The four-dimensional Fourier transform $\hat{\varphi}\left( q,z\right) $ of
the normalizable dual modes solves the reduced field equation%
\begin{equation}
\left[ \partial _{z}^{2}+\left( d-1\right) \left( a^{-1}\partial
_{z}a\right) \partial _{z}-\left( \partial _{z}\Phi \right) \partial
_{z}+q^{2}\right] \hat{\varphi}\left( q,z\right) =0  \label{fe}
\end{equation}%
obtained by variation of the bulk action (\ref{sd}). The corresponding
orthonormalized solutions $\psi _{n}\left( z\right) =N_{n}\hat{\varphi}%
\left( m_{n},z\right) $ have discrete momenta $q^{2}=m_{n}^{2}$ which
determine the glueball mass spectrum of the boundary gauge theory.

For dilaton fields which vanish at the UV boundary (as in the soft wall
case), the glueball correlator has the spectral representation 
\begin{eqnarray}
\hat{\Pi}\left( -q^{2}\right) &=&-i\int \frac{d^{4}q}{\left( 2\pi \right)
^{4}}e^{iq\left( x-y\right) }\left\langle T\mathcal{O}_{S}\left( x\right) 
\mathcal{O}_{S}\left( y\right) \right\rangle  \notag \\
&=&-\left( \frac{R^{3}}{\kappa \varepsilon ^{3}}\right) ^{2}\sum_{n}\frac{%
\psi _{n}^{\prime }\left( \varepsilon \right) \psi _{n}^{\prime }\left(
\varepsilon \right) }{q^{2}-m_{n}^{2}+i\bar{\varepsilon}}=-\sum_{n}\frac{%
f_{n}^{2}m_{n}^{4}}{q^{2}-m_{n}^{2}+i\bar{\varepsilon}}  \label{specrep}
\end{eqnarray}%
where a prime denotes differentiation with respect to $z$ and regularizing
contact terms for $\varepsilon \rightarrow 0$ are not written explicitly.
The pole residues of Eq. (\ref{specrep}) contain the decay constants 
\begin{equation}
f_{n}:=\frac{1}{m_{n}^{2}}\left\langle 0\left| \mathcal{O}_{S}\left(
0\right) \right| 0_{n}^{++}\right\rangle =\frac{R^{3}}{\kappa m_{n}^{2}}%
\frac{\psi _{n}^{\prime }\left( \varepsilon \right) }{\varepsilon ^{3}}
\label{dc}
\end{equation}%
of the $n$-th $0^{++}$ glueball excitation. Since the $f_{n}$ can be
regarded as the glueball (Bethe-Salpeter) wave functions at the origin, a
smaller glueball size implies a higher concentration of the wave function
and consequently a larger value of $f_{n}$. Evidence for such an enhancement
of the ground-state decay constant was found in instanton vacuum models \cite%
{sch95}, in QCD sum rule analyses which include instanton contributions to
the OPE coefficients \cite{for01}, and in (quenched) lattice simulations %
\cite{che06}.

\subsection{Hard wall}

The hard-wall geometry of Polchinski and Strassler \cite{pol02} is an AdS$%
_{5}$ slice with a Randall-Sundrum type cutoff $z_{m}$ at the IR brane, 
\begin{equation}
e^{2A^{\left( \text{hw}\right) }\left( z\right) }=\theta \left(
z_{m}-z\right) ,\,\ \ \ \ \ z_{m}\simeq \Lambda _{\text{QCD}}^{-1},\text{ \
\ \ \ \ }\Phi ^{\left( \text{hw}\right) }\equiv 0,  \label{hw}
\end{equation}%
which implements conformal symmetry in the UV and its breaking in the IR in
a minimal fashion.

The glueball decay constants in the hard-wall background can be calculated
directly from the normalized solutions%
\begin{equation}
\psi _{n}\left( z\right) =N_{n}\left( m_{n}z\right) ^{2}J_{2}\left(
m_{n}z\right)  \label{hwm}
\end{equation}%
(where $n=1,2,...$) of the field equation (\ref{fe}) in the metric (\ref{hw}%
). The constants $N_{n}$ are determined by the normalization condition $%
\int_{0}^{z_{m}}dz\left( R/z\right) ^{3}\psi _{n}^{2}=1$. For the Dirichlet
IR boundary condition $\psi _{n}\left( z_{m}\right) =0$ one then obtains %
\cite{bos03} 
\begin{equation}
m_{n}^{\left( \text{D}\right) }=\frac{j_{2,n}}{z_{m}},\text{ \ \ \ \ \ }%
N_{n}^{\left( \text{D}\right) }=\frac{\sqrt{2}}{m_{n}^{\left( \text{D}%
\right) 2}R^{3/2}z_{m}\left| J_{1}\left( j_{2,n}\right) \right| }
\label{mND}
\end{equation}%
while the alternative Neumann\ boundary condition $\psi _{n}^{\prime }\left(
z_{m}\right) =0$ yields \cite{bos03} 
\begin{equation}
m_{n}^{\left( \text{N}\right) }=\frac{j_{1,n}}{z_{m}},\text{ \ \ \ \ \ }%
N_{n}^{\left( \text{N}\right) }=\frac{\sqrt{2}}{m_{n}^{\left( \text{N}%
\right) 2}R^{3/2}z_{m}\left| J_{0}\left( j_{1,n}\right) \right| }.
\label{mNN}
\end{equation}%
Here $j_{m,n}$ denotes the $n$-th zero of the $m$-th Bessel function \cite%
{abr72}.

From the general expression (\ref{dc}) for the decay constants and the
hard-wall eigenmodes (\ref{hwm})\ one then finds \cite{for08}%
\begin{equation}
f_{n}=\lim_{\varepsilon \rightarrow 0}\frac{R^{3}}{\kappa m_{n}^{2}}\frac{%
\psi _{n}^{\prime }\left( \varepsilon \right) }{\varepsilon ^{3}}=\frac{N_{n}%
}{2}\frac{R^{3}}{\kappa }m_{n}^{2}
\end{equation}%
or more specifically for the above two IR boundary conditions%
\begin{equation}
f_{n}^{\left( \text{D}\right) }=\frac{1}{\sqrt{2}\left| J_{1}\left(
j_{2,n}\right) \right| }\frac{R^{3/2}}{\kappa z_{m}},\text{ \ \ \ \ \ }%
f_{n}^{\left( \text{N}\right) }=\frac{1}{\sqrt{2}\left| J_{0}\left(
j_{1,n}\right) \right| }\frac{R^{3/2}}{\kappa z_{m}}.  \label{fhw}
\end{equation}%
For a quantitative estimate one can fix the overall normalization factor $%
R^{3/2}/\kappa $ according to Eq. (\ref{est}) and set the IR scale $%
z_{m}^{-1}\sim \Lambda _{\text{QCD}}$ e.g. such that a typical quenched
ground-state glueball mass of $m_{S}\sim $ 1.5 GeV is reproduced, or at the
value $z_{m}^{-1}\simeq 0.35$ GeV found in the classical hadron sector.
Either way, the ground-state decay constant predictions remain in the range $%
f_{S}^{\left( \text{hw}\right) }\equiv f_{1}^{\left( \text{hw}\right)
}\simeq 0.8-0.9$ GeV \cite{for08}.

\subsection{Soft wall}

The hard-wall predictions (\ref{mND}), (\ref{mNN}) for the squared masses of
scalar glueballs (and of other hadrons)\ grow quadratically with high radial
(and orbital) excitation quantum numbers, in contrast to the linear
trajectories expected from semiclassical models and data. The dilaton
soft-wall background \cite{kar06}%
\begin{equation}
A^{\left( \text{sw}\right) }\left( z\right) \equiv 0,\text{ \ \ \ \ \ }\Phi
^{\left( \text{sw}\right) }\left( z\right) =\lambda ^{2}z^{2}  \label{sw}
\end{equation}%
provides an economical corrective to this problem in the meson \cite{kar06}
and glueball \cite{col207} sectors. (The ``metric soft wall'' \cite{for07}
is a dilaton-less alternative which also yields linear baryon mass
trajectories.)

The spectrum-generating normalizable solutions of Eq. (\ref{fe}) in the
background (\ref{sw}) are those Kummer functions whose power series
expansion truncates to generalized Laguerre polynomials $L_{n}^{\left(
2\right) }$ \cite{abr72}, i.e.%
\begin{equation}
\psi _{n}\left( z\right) =N_{n}\lambda ^{4}z^{4}{}_{1}F_{1}\left(
-n,3,z^{2}\lambda ^{2}\right) =N_{n}\lambda ^{4}z^{4}\frac{n!}{\left(
3\right) _{n}}L_{n}^{\left( 2\right) }\left( \lambda ^{2}z^{2}\right)
\label{swnm}
\end{equation}%
where $n=0,1,2,...$, $\left( a\right) _{n}\equiv a\left( a+1\right)
...\left( a+n-1\right) $ and $_{1}F_{1}$ is a confluent hypergeometric
function \cite{abr72}. The ensuing restriction to discrete $q^{2}=m_{n}^{2}$
generates the mass gap $m_{0}=2\sqrt{2}\lambda $ and the glueball mass
spectrum \cite{col207}%
\begin{equation}
m_{n}^{2}=4\left( n+2\right) \lambda ^{2}  \label{msw}
\end{equation}%
which indeed lies on a linear Pomeron-type trajectory. The condition $%
\int_{0}^{\infty }dz\left( R/z\right) ^{3}\exp \left( -\lambda
^{2}z^{2}\right) \psi _{n}^{2}\left( z\right) =1$ fixes the normalization
constants%
\begin{equation}
N_{n}=\lambda ^{-1}R^{-3/2}\sqrt{\frac{\left( n+1\right) \left( n+2\right) }{%
2}}\overset{n\gg 3}{\longrightarrow }2^{-1/2}\lambda ^{-1}R^{-3/2}n.
\end{equation}

From the general expression (\ref{dc}) one then finds the glueball decay
constants in the soft-wall background as \cite{for08} 
\begin{equation}
f_{n}=2\sqrt{2\left( n+1\right) \left( n+2\right) }\frac{\lambda ^{3}R^{3/2}%
}{m_{n}^{2}\kappa }=\frac{1}{\sqrt{2}}\sqrt{\frac{n+1}{n+2}}\frac{\lambda
R^{3/2}}{\kappa }.{}  \label{fsw}
\end{equation}%
After fixing the factor $R^{3/2}/\kappa $ by Eq. (\ref{est}) one can as
above obtain quantitative estimates for the $f_{n}$ by setting the IR scale $%
\lambda $ either such as to reproduce the typical quenched mass value $%
m_{S}\sim $ 1.5 GeV or by using the value $\lambda \simeq \sqrt{2}\Lambda _{%
\text{QCD}}\simeq 0.49$ GeV of Ref. [18]. Both variants lead to similar
soft-wall predictions $f_{S}^{\left( \text{sw}\right) }\equiv f_{0}^{\left( 
\text{sw}\right) }\simeq 0.3$ GeV for the ground state decay constant \cite%
{for08}.

\section{Holographic glueball correlators}

Holographic correlation functions can be derived from the on-shell action of
the gravity dual which plays the role of their generating functional. To
construct it, one employs the bulk-to-boundary propagator $\hat{K}\left(
q,z\right) $ \cite{wit98}, i.e. the solution of the field equation (\ref{fe}%
) subject to the $\hat{K}\left( q;\varepsilon \rightarrow 0\right) =1$ and $%
\hat{K}\left( 0;z\right) =1$ boundary conditions, to write the solution of
Eq. (\ref{fe}) with a boundary source $\varphi ^{\left( s\right) }\left(
x^{\prime }\right) $ as 
\begin{equation}
\varphi \left( x,z\right) =\int \frac{d^{4}q}{\left( 2\pi \right) ^{4}}%
e^{-iqx}\hat{K}\left( q,z\right) \int d^{4}x^{\prime }e^{iqx^{\prime
}}\varphi ^{\left( s\right) }\left( x^{\prime }\right) .  \label{sln}
\end{equation}%
Inserting the solution (\ref{sln}) into Eq. (\ref{sd}) yields the on-shell
action, and taking two functional derivatives with respect to $\varphi
^{\left( s\right) }$ then generates the correlator%
\begin{equation}
\hat{\Pi}\left( -q^{2}\right) =-\frac{R^{3}}{\kappa ^{2}}\left[ \frac{%
e^{-\Phi \left( z\right) }}{z^{3}}\hat{K}\left( q,z\right) \partial _{z}\hat{%
K}\left( q,z\right) \right] _{z=\varepsilon \rightarrow 0}  \label{corr}
\end{equation}%
of the scalar glueball. Analytical solutions for $\hat{K}\left( q,z\right) $
in both hard- and soft-wall backgrounds were found in Ref. [6].

\subsection{Hard wall}

After plugging the analytical hard-wall solution for $\hat{K}\left(
q,z\right) $ \cite{for08} (subject to the Neumann IR boundary condition)
into the general expression (\ref{corr}) and discarding two contact terms,
one ends up with the correlator \cite{for08}%
\begin{equation}
\hat{\Pi}\left( Q^{2}\right) =\frac{R^{3}}{8\kappa ^{2}}Q^{4}\left[ 2\frac{%
K_{1}\left( Qz_{m}\right) }{I_{1}\left( Qz_{m}\right) }-\ln \left( \frac{%
Q^{2}}{\mu ^{2}}\right) \right]  \label{hwc}
\end{equation}%
($K_{\nu },I_{\nu }$ are McDonald functions \cite{abr72}) at spacelike
momenta $Q^{2}=-q^{2}$. Its spectral density $\rho \left( s\right) $ is
defined by means of the dispersion relation%
\begin{equation}
\hat{\Pi}\left( Q^{2}\right) =\int_{m_{\text{min}}^{2}}^{\infty }ds\frac{%
\rho \left( s\right) }{s+Q^{2}}  \label{drel}
\end{equation}%
(suppressing again subtraction terms)\ and takes the form \cite{for08}%
\begin{equation}
\rho \left( s\right) =\frac{R^{3}}{2\kappa ^{2}z_{m}^{2}}s^{2}\sum_{n=1}^{%
\infty }\frac{\delta \left( s-m_{n}^{2}\right) }{J_{0}^{2}\left(
j_{1,n}\right) }  \label{hwsd}
\end{equation}%
where the\ hard-wall mass spectrum $m_{n}=$ $j_{1,n}/z_{m}$ reappears. The
spectral weight (\ref{hwsd}) is non-negative and consists of zero-width
poles, as expected in the large-$N_{c}$ limit where glueballs are stable
against strong decay.

The overall correlator normalization $R^{3}/\kappa ^{2}$ is fixed by
matching the leading conformal logarithm to the free QCD gluon loop,%
\begin{equation}
\frac{R^{3}}{\kappa ^{2}}=\frac{2\left( N_{c}^{2}-1\right) }{\pi ^{2}}
\label{est}
\end{equation}%
with $N_{c}=3$. For $Q\gg \mu >z_{m}^{-1}$ the holographic correlator (\ref%
{hwc}) can be compared to the QCD short-distance expansion \cite{for01}. The
exponential $Q^{2}$ dependence (times powers of $Q^{2}$) of its
non-conformal part%
\begin{equation}
\hat{\Pi}^{\left( \text{np}\right) }\left( Q^{2}\right) \overset{Qz_{m}\gg 1}%
{\longrightarrow }\frac{4}{\pi }\left[ 1+\frac{3}{4}\frac{1}{Qz_{m}}+O\left( 
\frac{1}{\left( Qz_{m}\right) ^{2}}\right) \right] Q^{4}e^{-2Qz_{m}}
\label{nph}
\end{equation}%
has its QCD OPE counterpart in the small-size instanton contribution \cite%
{for01} \textbf{\ }%
\begin{equation}
\hat{\Pi}^{\left( I+\bar{I}\right) }\left( Q^{2}\right) \overset{Q\bar{\rho}%
\gg 1}{\longrightarrow }2^{4}5^{2}\pi \zeta \bar{n}\left( Q\bar{\rho}\right)
^{3}e^{-2Q\bar{\rho}}  \label{npi}
\end{equation}%
to the unit operator coefficient. Since the instanton-induced correlations
are attractive and of relatively short range $\sim \bar{\rho}$, they reduce
the scalar glueball mass and size while increasing its decay constant \cite%
{for01}.

Approximately equating Eq. (\ref{npi}) to the second term in Eq. (\ref{nph})
yields the holographic estimates \textbf{\ }%
\begin{equation}
\bar{\rho}\simeq z_{m}\text{, \ \ \ \ \ \ }\bar{n}\simeq \frac{3}{%
2^{4}5^{2}\pi ^{2}\zeta }\frac{1}{z_{m}^{4}}  \label{holest}
\end{equation}%
for the average instanton size $\bar{\rho}$ and the overall instanton
density $\bar{n}$ in terms of the IR scale $z_{m}$. The first relation
reflects the duality between gauge instantons of size $\rho $ and pointlike $%
D$ instantons localized at $z=\rho $. With $z_{m}^{-1}\sim \Lambda _{\text{%
QCD}}\simeq 0.33$ GeV it implies $\bar{\rho}\sim 0.6$ fm, i.e. almost twice
the standard\ value \cite{sch95}. This may suggest that the bulk dynamics (%
\ref{sd}) is more suitable for pure Yang-Mills theory with it larger
instanton sizes $\bar{\rho}\simeq 0.4-0.5$ fm, and that the strongly coupled
hard-wall UV dynamics describes the small-instanton physics beyond the
conformal regime rather poorly.

The above discussion implies that $\rho ^{\left( \text{hw}\right) }\leq
z_{m}\sim \mu ^{-1}$, i.e. that large instantons, which would contribute to
the condensates, are absent since their duals do not fit into the AdS$_{5}$
slice. Hence the hard wall reproduces the QCD result that the \emph{hard}
nonperturbative physics from small instantons (instead of the soft
condensate physics) dominates the short-distance $0^{++}$ glueball
correlator \cite{for01}. Moreover, the QCD low-energy theorem \cite{let} 
\begin{equation}
\hat{\Pi}\left( 0\right) =\frac{32\pi }{\alpha _{s}b_{0}}\left\langle
G^{2}\right\rangle +O\left( m_{q}\right)  \label{let}
\end{equation}%
with $b_{0}=11N_{c}/3-2N_{f}/3$ is trivially satisfied by the hard-wall
correlator (\ref{hwc}) which vanishes at $Q^{2}=0$. This is consistent with
Eq. (\ref{let}) since the gluon condensate vanishes in the hard wall as well.

\subsection{Soft wall}

Inserting the analytical solution for the soft-wall bulk-to-boundary
propagator $\hat{K}\left( q;z\right) $ \cite{for08} into Eq. (\ref{corr})
yields (after discarding two divergent contact terms) the soft-wall
correlator \cite{for08}%
\begin{equation}
\hat{\Pi}\left( Q^{2}\right) =-\frac{2R^{3}}{\kappa ^{2}}\lambda ^{4}\left[
1+\frac{Q^{2}}{4\lambda ^{2}}\left( 1+\frac{Q^{2}}{4\lambda ^{2}}\right)
\psi \left( \frac{Q^{2}}{4\lambda ^{2}}\right) \right]  \label{swc}
\end{equation}%
in terms of the digamma function $\psi \left( z\right) =\Gamma ^{\prime
}\left( z\right) /\Gamma \left( z\right) $ \cite{abr72}. The analyticity
structure of Eq. (\ref{swc}) implies that its spectral density has the form %
\cite{for08}%
\begin{equation}
\rho \left( s\right) =\frac{\lambda ^{2}R^{3}}{2\kappa ^{2}}s\left(
s-m_{0}^{2}/2\right) \sum_{n=0}^{\infty }\delta \left( s-m_{n}^{2}\right)
=\sum_{n=0}^{\infty }f_{n}^{2}m_{n}^{4}\delta \left( s-m_{n}^{2}\right)
\label{swsd}
\end{equation}%
which is non-negative for $s\geq m_{0}^{2}/2$ and consists of zero-width
poles, as expected at large $N_{c}$, at the soft-wall masses (\ref{msw})
with residua determined by the soft-wall decay constants (\ref{fsw}).

In order to compare the soft-wall correlator (\ref{swc}) to the QCD OPE, we
use the asymptotic expansion of the digamma function to rewrite Eq. (\ref%
{swc}) for $Q^{2}\gg 4\lambda ^{2}>\Lambda _{\text{QCD}}^{2}$ as 
\begin{equation}
\hat{\Pi}\left( Q^{2}\right) =-\frac{2}{\pi ^{2}}Q^{4}\left[ \ln \frac{Q^{2}%
}{\mu ^{2}}+\frac{4\lambda ^{2}}{Q^{2}}\ln \frac{Q^{2}}{\mu ^{2}}+\frac{%
2^{2}5}{3}\frac{\lambda ^{4}}{Q^{4}}-\frac{2^{4}}{3}\frac{\lambda ^{6}}{Q^{6}%
}+\frac{2^{5}}{15}\frac{\lambda ^{8}}{Q^{8}}+...\right]  \label{c}
\end{equation}%
which is renormalized at the OPE scale $\mu $. The normalization $%
R^{3}/\kappa ^{2}$ is again fixed by Eq. (\ref{est}) since large momenta $Q$
probe the $z\rightarrow 0$ region where hard- and soft-wall correlators are
governed by the same AdS$_{5}$-induced logarithm.

Besides the leading conformal and a second logarithmic term, the expansion (%
\ref{c}) contains an infinite tower of power corrections. Comparison with
the OPE suggests those to be related to the gauge-theory condensates $%
\left\langle \mathcal{O}_{D}\right\rangle \sim \lambda ^{D}$ of $D=4,6,8,...$
dimensional composite operators. For a first order-of-magnitude check one
may equate the coefficients of the $D=4,6$ and $8$ terms in Eq. (\ref{c}) to
the ($O\left( \alpha _{s}^{0}\right) $) QCD Wilson coefficients, yielding%
\begin{equation}
\left\langle G^{2}\right\rangle \simeq -\frac{10}{3\pi ^{2}}\lambda ^{4},%
\text{ \ \ \ }\left\langle gG^{3}\right\rangle \simeq \frac{4}{3\pi ^{2}}%
\lambda ^{6},\text{ \ \ \ }\left\langle G^{4}\right\rangle \simeq -\frac{8}{%
15\pi ^{3}\alpha _{s}}\lambda ^{8}.  \label{g4}
\end{equation}%
For $\lambda \sim \Lambda _{\text{QCD}}$ these are the rough magnitudes of
the QCD condensates, but the sign of the QCD gluon condensate $\left\langle
G^{2}\right\rangle \sim 0.4-1.2$ GeV$^{4}$ is positive. While QCD estimates
of both signs exist for the three-gluon condensate, the above signs would
also be at odds with the factorization approximation $\left\langle
G^{4}\right\rangle \simeq \left( 9/16\right) \langle G^{2}\rangle ^{2}$.

The probably most intriguing prediction of the soft-wall correlator (\ref{c}%
) is the additional power correction of dimension two which cannot appear in
the OPE since QCD lacks a corresponding local operator. When linear
contributions to the \emph{short}-distance heavy-quark potential are
approximately described by a tachyonic gluon mass $\bar{\lambda}$, however,
one finds the correction \cite{che99}%
\begin{equation}
\hat{\Pi}_{\bar{\lambda}}^{(\text{CNZ})}\left( Q^{2}\right) =-\frac{12}{\pi
^{2}}\bar{\lambda}^{2}Q^{2}\ln \frac{Q^{2}}{\mu ^{2}}  \label{cnz}
\end{equation}%
which has precisely the form of the second term in Eq. (\ref{c}). Comparison
of the coefficients provides the holographic estimate $\bar{\lambda}%
^{2}\simeq \left( 2/3\right) \lambda ^{2}$ and with $\lambda \simeq \sqrt{2}%
\Lambda _{\text{QCD}}$ further $\bar{\lambda}^{2}\simeq 0.15$ GeV$^{2}$
which is of the expected magnitude \cite{che99} but again of opposite sign,
i.e. not tachyonic.

Hence the expansion (\ref{c}) completely reproduces the qualitative $Q^{2}$
dependence of the QCD short-distance correlator (to leading order in $\alpha
_{s}$) but fails to predict the sign of at least the two leading terms. This
pattern can be better understood by recalling that the dimensions of the QCD
condensates are generated by operators which are renormalized at $\mu
\lesssim 1$ GeV and thus IR\ dominated. The form and general $Q^{2}$
dependence of the QCD power corrections is therefore governed by IR physics,
which may explain why the strongly-coupled soft-wall dynamics can reproduce
it. The deviations of size and signs of the holographic power corrections
from their QCD counterparts (and the absence of radiative corrections)
should then originate mainly from the poorer description of the
weakly-coupled, perturbative Wilson coefficients by the soft-wall dynamics
which lacks $\alpha ^{\prime }$ corrections and remains strongly coupled in
the UV.

The above interpretation also suggests an approximate separation of the
soft-wall power corrections into Wilson coefficients and condensates.
Indeed, under the premise that the strongly-coupled soft wall dynamics
approximately reproduces the QCD\ condensate values, one may obtain
holographic estimates for the Wilson coefficients. The gluon condensate
coefficient, e.g., becomes with $\left\langle G^{2}\right\rangle \simeq
\left( 20/3\right) \Lambda _{\text{QCD}}^{4}$ \cite{for01} and $\lambda
\simeq \sqrt{2}\Lambda _{\text{QCD}}$ \cite{for07} 
\begin{equation}
C_{\left\langle G^{2}\right\rangle }^{\left( \text{sw}\right) }\simeq -\frac{%
8}{\pi ^{2}}=-\frac{2}{\pi ^{2}}C_{\left\langle G^{2}\right\rangle }^{\left( 
\text{QCD,lo}\right) }.  \label{CG2}
\end{equation}%
Its smaller size and opposite sign relative to the QCD result provides some
intuition for the soft-wall deficiencies in describing weakly-coupled QCD
physics. Of course, the estimate (\ref{CG2}) is prone to further error
sources, including uncertainties in the QCD condensate values and their
sensitivity to light quark contributions. (The above separation into hard
and soft contributions would fail for the two-dimensional power correction,
incidentally, since both the gluon mass $\bar{\lambda}$ and its coefficient
receive UV contributions.) Since the condensates are hadron-channel
independent while the Wilson coefficients are not, one would further expect
to obtain inconsistent condensate estimates when trying to extract them in
different channels by relying on the respective QCD Wilson coefficients.
Finally, we note that the soft-wall correlator (\ref{swc}) with $\hat{\Pi}%
\left( 0\right) =0$ violates the low-energy theorem (\ref{let}) since Eq. (%
\ref{g4}) implies a finite RHS.

\subsection{A generalized soft-wall correlator?}

Recently an attempt has been made to generalize the soft-wall glueball
correlator (\ref{swc}) by adding to the bulk-to-boundary propagator $\hat{K}%
\left( q,z\right) $ the at small $z$ subleading solution of Eq. (\ref{fe}),
multiplied by an a priori arbitrary coefficient function $\tilde{B}\left(
Q^{2}\right) $ \cite{col07}. The added solution blows up at large $z$, i.e.
it violates the standard ``regularity in the bulk'' condition \cite{wit98}
and requires an ad-hoc IR cutoff prescription without obvious correspondence
on the gauge-theory side (in contrast to the standard UV renormalization
whose ``dual'' tames the volume divergence of the on-shell bulk action). The
resulting expression for the correlator differs from ours, i.e. Eq. (\ref%
{swc}), by the addition of the arbitrary coefficient function $\tilde{B}%
\left( Q^{2}\right) $. Any desired behavior of the correlator could thus be
chosen by hand, and independently of the soft-wall background, by adapting $%
\tilde{B}\left( Q^{2}\right) $ accordingly. (Ref. [12] attempted to use this
apparent freedom to equate the $D\geq 4$ power corrections to their QCD\
values and to eliminate the two-dimensional power correction.)

Hence the prescription of Ref. [12] gives up on the one-to-one duality
between the gauge vacuum and the gravity background (together with other
parts of the AdS/CFT dictionary) and results in a practically total loss of
predictive power. Moreover, it is internally inconsistent. Indeed, it was
shown to result in the same mass spectrum (\ref{msw}) and the same decay
constants (\ref{fsw}) as in our case \cite{col07}. Hence it must reproduce
our spectral density (\ref{swsd}) and thus the physics content of the
correlator (\ref{swc}), i.e. any remaining discrepancies have to originate
from the unphysical contact terms which are needed to regularize the
dispersion integral (\ref{drel}). This is in direct contradiction to the
supposed freedom of adding an arbitrary function $\tilde{B}\left(
Q^{2}\right) $ to the correlator.

The loss of predictivity, uniqueness and consistency incurred when relaxing
the regularity of $\hat{K}\left( q,z\right) $ in the bulk has a common
origin. Indeed, any given (smooth) function defined on the UV boundary $%
S^{d} $ of (Euclidean and compactified) AdS$_{d+1}$ is known to have a \emph{%
unique} extension to a solution of the massless scalar field equation in the
AdS$_{d+1}$ bulk \cite{wit98}. When applied to the boundary source $\varphi
^{\left( s\right) }\left( x\right) $, this mathematical fact ensures the
one-to-one correspondence between the gauge-theory operators and the dual
string mode solutions. The inconsistency of the attempted generalization of
the bulk-to-boundary propagator arises from the violation of this fact. In
order to maintain the mathematically required uniqueness of the relation (%
\ref{sln}) between a given boundary source $\varphi ^{\left( s\right)
}\left( x\right) $ and its dual mode solution $\varphi \left( x,z\right) $,
the UV-subleading solution must not be added to $\hat{K}\left( q,z\right) $.

\section{Summary and conclusions}

We have derived and analyzed the predictions of the two currently most
popular AdS/QCD duals, i.e. the hard-wall and dilaton soft-wall backgrounds,
for the $0^{++}$ glueball correlation function and decay constants.

In their representation of specific nonperturbative glueball physics (at
momenta larger than the QCD scale) both holographic duals turn out to
complement each other: the soft-wall correlator contains all known types of
QCD power corrections, generated both by vacuum condensates and by a
hypothetical UV gluon mass, while sizeable exponential corrections of the
type induced by small-scale QCD instantons are reproduced in the hard-wall
correlator. Since the QCD power corrections to the $0^{++}$ glueball
correlator are suppressed by unusually small Wilson coefficients whereas the
small-instanton contributions are enhanced, the hard-wall background may
provide a more reliable approximation to the scalar glueball correlator.
Furthermore, the above complementarity, which helps to relate holographic
predictions (and their limitations) to specific aspects of the gauge
dynamics, should extend to other hadron channels.

While all our holographic estimates have the order of magnitude expected
from QCD$,$ the signs of the two leading soft-wall power corrections are
opposite to those of standard QCD estimates (and in conflict with the
factorization approximation for the four-gluon condensate). We have argued
that this provides evidence for the short-distance physics in the Wilson
coefficients to be inadequately reproduced by the strongly-coupled UV
dynamics of bottom-up models (beyond the leading conformal logarithm). We
have further shown that this problem cannot be mended by admixing the
UV-subleading solution to the bulk-to-boundary propagator (as recently
advocated)\ without loosing consistency and predictive power.

In addition, we have provided first holographic estimates for the $0^{++}$
glueball decay constants which contain glueball size information, are
important for experimental glueball searches and probe aspects of the dual
dynamics to which the mass spectrum is less sensitive. The hard-wall
prediction for the ground-state decay constant $f_{S}$ is more than twice as
large as its soft-wall counterpart. This is a consequence of the exponential
contributions to the hard-wall correlator which reproduce the strong
instanton-induced short-distance attraction in the QCD correlator. The
hard-wall prediction $f_{S}^{\left( \text{hw}\right) }\simeq 0.8-0.9$ GeV
implies an exceptionally small glueball size and agrees inside errors with
IOPE sum-rule and lattice results.

We would like to thank Pietro Colangelo and his coworkers for correspondence
and acknowledge financial support from the Funda\c{c}\~{a}o de Amparo a
Pesquisa do Estado de S\~{a}o Paulo (FAPESP) and the Deutsche
Forschungsgemeinschaft (DFG).


\begin{thebibliography}{99}
\bibitem{aha00} O. Aharony, S.S. Gubser, J.M. Maldacena, H. Ooguri and Y.
Oz, Phys. Rep. \textbf{323}, 183\ (2000).

\bibitem{revs} K. Peeters and M. Zamaklar, Eur. Phys. J. Special Topics 
\textbf{152}, 113 (2007); S.J. Brodsky and G.F. de T\'{e}ramond,
arXiv:0802.0514; J. Erdmenger, N. Evans, I. Kirsch and E. Threlfall, Eur.
Phys. J. A \textbf{35}, 81 (2008).

\bibitem{pol02} J. Polchinski and M. J. Strassler, Phys. Rev. Lett. \textbf{%
88}, 031601 (2002).

\bibitem{kar06} A. Karch, E. Katz, D.T. Son and M.A. Stephanov, Phys. Rev. D 
\textbf{74}, 015005 (2006).

\bibitem{ff} H.R. Grigoryan and A.V. Radyushkin, Phys. Lett. B \textbf{650},
421 (2007); Phys. Rev. D \textbf{76}, 095007 (2007); Phys. Rev. D \textbf{76}%
, 115007 (2007); Phys. Rev. D \textbf{77}, 115024 (2008); S.J. Brodsky and
G.F. de T\'{e}ramond, Phys. Rev. D. \textbf{77}, 056007 (2008);
arXiv:0804.0452; H.J. Kwee and R.F. Lebed, JHEP \textbf{0801}, 027 (2008);
D.K. Hong, M. Rho, H.-U. Yee and P. Yi, Phys. Rev. D \textbf{77}, 014030
(2008); K. Hashimoto, T. Sakai and S. Sugimoto, arXiv:0806.3122; K.Y. Kim
and I. Zahed, arXiv:0807.0033.

\bibitem{for08} H. Forkel, Phys. Rev. D \textbf{78}, 025001 (2008).

\bibitem{che06} Y. Chen et al., Phys. Rev. D \ \textbf{73}, 014516\ (2006);
M. Loan and Y. Ying, hep-lat/0603030; N. Ishii, H. Suganuma and H.
Matsufuru, Phys. Rev. D \textbf{66}, 94506 (2002); P. de Forcrand and K.-F.
Liu, Phys. Rev. Lett. \textbf{69}, 245 (1992); R. Gupta et al., Phys. Rev. D 
\textbf{43}, 2301 (1991).

\bibitem{for01} H. Forkel, Phys. Rev. D \textbf{64}, 034015 (2001); Phys.
Rev. D \textbf{71}, 054008\ (2005); Proceedings of ``Continuous advances in
QCD'', Minneapolis (2006), 383 [arXiv:hep-ph/0608071].

\bibitem{che99} K.G. Chetyrkin, S. Narison and V.I. Zakharov, Nucl. Phys. 
\textbf{B 550}, 353\ (1999).

\bibitem{let} M.A. Shifman, Phys. Rep. \textbf{209}, 341 (1991).

\bibitem{sch08} T. Sch\"{a}fer, Phys. Rev. D \textbf{77}, 126010 (2008).

\bibitem{col07} P. Colangelo, F. De Fazio, F. Jugeau and S. Nicotri,
arXiv:0711.4747.

\bibitem{zuo08} F. Zuo and T. Huang, arXiv:0801.1172.

\bibitem{sch95} T. Sch\"{a}fer and E.V. Shuryak, Phys. Rev. Lett. \textbf{75}%
, 1707 (1995).

\bibitem{bos03} H.~Boschi-Filho and N.~R.~F.~Braga, JHEP \textbf{05}, 009
(2003); H.~Boschi-Filho, N.~R.~F.~Braga and H.~L.~Carrion,\ Rev.\ D\textbf{\
73}, 047901 (2006).

\bibitem{abr72} M. Abramowitz and I.A. Stegun, \textit{Handbook of
Mathematical Functions} (U.S. GPO, Washington, DC, 1972).

\bibitem{col207} P. Colangelo, F. De Fazio, F. Jugeau and S. Nicotri, Phys.
Lett. \textbf{B 652}, 73\ (2007).

\bibitem{for07} H. Forkel, T. Frederico and M. Beyer, JHEP \textbf{07}, 077
(2007); Int. J. Mod. Phys. E \textbf{16}, 2794\ (2007) [arXiv:0705.4115].

\bibitem{wit98} E.~Witten, Adv.\ Theor.\ Math.\ Phys.\ \textbf{2}, 253
(1998).
\end{thebibliography}
\end{document}